\documentclass[aps,prl,twocolumn,showpacs,floatfix,nofootinbib,superscriptaddress]{revtex4}

\usepackage{amssymb,graphics,graphicx,times,bm,bbm,multirow,color}

\begin{document}

\title{Optimal control of a leaking qubit}
\author{P.~Rebentrost}\email{pr@patrickre.com}
\affiliation{Department of Chemistry and Chemical Biology,
Harvard University, 12 Oxford St., Cambridge, MA 02138, USA}
\affiliation{IQC and Department of Physics and Astronomy, University
of Waterloo, 200 University Ave W, Waterloo, ON, N2L 3G1, Canada}
\affiliation{Department Physik, Ludwig-Maximilians-Universit\"at,
Theresienstr. 37, 80333 Munich, Germany}
\author{F.K.~Wilhelm} \email{fwilhelm@iqc.ca}
\affiliation{IQC and Department of Physics and Astronomy, University
of Waterloo, 200 University Ave W, Waterloo, ON, N2L 3G1, Canada}
\date{\today}
\begin{abstract}
Physical implementations of quantum bits can contain coherent transitions to energetically close non-qubit states.
In particular, for anharmonic oscillator systems such as the superconducting phase qubit
and the transmon a two-level approximation is insufficient.
We apply optimal control theory to the envelope of a resonant Rabi pulse
in a qubit in the presence of a single, weakly off-resonant leakage level.
The gate error of a spin flip operation reduces by orders of magnitude compared to simple pulse shapes.
Near-perfect gates can be achieved for any pulse duration longer than an intrinsic limit given by the nonlinearity.
The pulses can be understood as composite sequences that refocus the leakage transition.
We also discuss ways to improve the pulse shapes.
\end{abstract}


\newcommand{\bra}[1]{\langle #1|}
\newcommand{\ket}[1]{|#1\rangle}
\newcommand{\braket}[2]{\langle #1|#2\rangle}

\maketitle

An ideal Hilbert space for a qubit has two dimensions,
spanned by the states $\ket{0}$ and $\ket{1}$. However, in physical implementations
quantum evolution can lead to transitions to non-qubit levels.
In most cases the levels are energetically far away and hence are not the dominating limitation
for quantum operations,  for instance in the case of orbital degrees of freedom in nuclear or electronic spin-$1/2$ qubits
\cite{NielsenBook}.
In other cases such as in superconducting qubits \cite{Insight} and optical lattices \cite{Maneshi08}
the situation is less favorable. In superconducting charge qubits the tunneling of more than one excess Cooper-pair
on the superconducting island still has a sizable energy penalty \cite{Makhlin01}.
In phase qubits \cite{Martinis02, Steffen06, Lucero08}, optical lattices, and
the transmon \cite{Koch07,Schreier08} one has a third level which is only slightly detuned
from the qubit energy splitting. These anharmonic oscillator qubits
are the motivation for the present paper. Control strategies are required that go beyond the
simplified two-level approximation. Initial phase qubit experiments
\cite{Martinis02} used weak and thus slow Rabi pulses for qubit manipulations,
making poor use of the available coherence time. Newer experiments
\cite{Lucero08} combine Gaussian pulses with engineered spectrum and
limited bandwidth. Initial proposals for avoiding leakage in
superconducting qubits based on renormalization do not apply to phase
qubits \cite{Fazio00} or lead to complex sequences of hard pulses
\cite{Tian00}.

Theoretical optimal control approaches are frequently used in quantum computing to synthesize quantum gates
\cite{Khaneja05,Palao02,Ramakrishna96}.
The models mostly assume coherent evolution of coupled two-level systems such as nuclear spins \cite{ColmRyanNMR}
or superconducting qubits \cite{Sporl07,Montangero07}.
Progress also has been made in optimally controlling systems in the presence of decoherence mechanisms,
either in Markovian \cite{Tosh06, Jirari06} or non-Markovian
\cite{Rebentrost06,Gordon08} regimes, as well as an inhomogeneity \cite{Mottonen06,Steffen07}.
In this paper, we study a qubit in the presence of coherent transitions to a weakly off-resonant non-qubit level.
We look for optimal pulses that generate quantum gates in the qubit
subspace, assuming that a {\em single} control for the envelope of a resonant pulse is available.
The pulse shapes significantly improve the fidelity obtained by
refined standard techniques \cite{Steffen03}. We identify a time limit related to the nonlinearity
above which efficient high fidelity pulses are possible.

{\it Model} $-$ Our Hamiltonian of a qubit in the presence of a single leakage level is given by:
\begin{equation}\label{eq:Hamiltonian}
H(t)=\epsilon \bar{\sigma}_z + \delta(t) \bar{\sigma}_x + E_{\rm L} \ket{L}\bra{L} +
\sqrt{2} \delta(t) (\ket{1}\bra{L} + {\rm h.c}).
\end{equation}
Here, $\epsilon$ is the qubit level splitting while $\delta(t)$ is a time dependent control field. The Pauli matrices for the qubit are given by $\bar{\sigma}_z=\vert 1 \rangle \langle 1 \vert - \vert 0 \rangle \langle 0 \vert$ and
$\bar{\sigma}_x=\vert 1 \rangle \langle 0 \vert + \vert 0 \rangle \langle 1 \vert$.
Non-zero controls generate transitions not only in the qubit subspace but also to the
leakage level $\ket{L}$ with energy $E_{\rm L}$. Hence, conventional control techniques can
reduce the quality of operations in the qubit subspace.
This Hamiltonian approximates the three lowest energy states of a weakly nonlinear oscillator
system, such as a Josephson phase qubit or a transmon \cite{Geller07, Steffen03}.
The matrix element of the transition $\ket{1} \to \ket{L}$ is scaled by $\sqrt{2}$.
The detuning of the leakage level is the difference between the qubit splitting and
the splitting of the transition $\ket{1} \to \ket{L}$, i.e.~$\Delta\omega=3\epsilon-E_{\rm L}$.
Matrix elements between $\ket{0}$ and $\ket{L}$ are negligible
following the parity selection rule of a nonlinear oscillator qubit
at weak to intermediate detuning.
For resonant driving on the qubit energy splitting,
$\delta(t)=\lambda(t)\cos(\omega t)$ with $\omega=2\epsilon$, one finds the Hamiltonian in the
rotating wave approximation (RWA),
\begin{equation}\label{eq:HamiltonianRwa}
H^{\rm R}(t)= -\Delta\omega \ket{L}^{\rm R}\bra{L}^{\rm R} + \lambda(t) \bar{\sigma}_x^{\rm R}
+ \sqrt{2} \lambda(t) (\ket{1}^{\rm R}\bra{L}^{\rm R} + {\rm h.c.}).
\end{equation}
A superscript $R$ denotes states and operators in the rotating frame, while
$\lambda(t)$ is the envelope of the resonant pulse and can be viewed as a modulated Rabi frequency.
The RWA representation assumes first that the logical qubits are encoded in
the rotating frame; we drop the superscript $R$ for notational clarity.
Second, it is assumed that the Rabi frequency $\lambda(t)$ is much smaller than $\epsilon$.
This assumption is validated by results of this paper that show, that the Rabi
frequency is ideally comparable with $\Delta\omega$.
In the following, we will naturally focus on the qubit flip operation (NOT gate) given
by,
\begin{equation}\label{eq:FinalGate}
U_{\rm F}=e^{i\varphi_1} \left ( e^{i\varphi_2} \ket{L}\bra{L} + \bar{\sigma}_x \right ).
\end{equation}
The global phase and
the relative phase $\phi_1$ and the phase
of the leakage level $\phi_2$ are meaningless for the NOT gate and
thus
taken to be arbitrary.
First, we discuss controllability in the limits of weak ($\lambda\ll \Delta\omega$) and
strong ($\lambda\gg \Delta\omega$) driving: for weak driving it is possible to
make the population of the leakage level arbitrarily small by choosing
the $\lambda$ small enough.
The downside of this method is that the pulse has to be very long, therefore gate fidelity is limited by the
coherence time.
In the limit of strong driving, one in principle could make the pulses
arbitrarily short and apply many of them within the coherence
time. However, the strong driving Hamiltonian, in which $\Delta\omega$
is the smallest frequency, is similar to that of a
single resonantly driven harmonic oscillator,
i.e.~$H(t)\approx \lambda(t) \sigma_x + \sqrt{2} \lambda(t) (\ket{1}\bra{L} + \ket{L}\bra{1})$.
Hence, no high-amplitude control field creates a perfect NOT in the qubit subspace.

{\it Approximate solution} $-$ One can find an approximate solution for our problem
in the weak driving limit, $\lambda/\Delta\omega \ll 1$, that is exact
to first order in $\lambda/\Delta\omega$ \cite{Steffen03}.
We denote by $W(t)$ a free evolution ($\lambda=0$) of time $t$ and by
$R(\theta)$ a weakly driven evolution, such that $\theta=\int_0^t \lambda(t') dt'$ with $\lambda/\Delta\omega \ll 1$.
Then the solution is given by
\begin{equation}\label{eq:RotationQubit}
U_{\rm F} \simeq R(\theta_1) W\left(\frac{\pi}{\Delta\omega}\right) R(\theta_2)
W\left(\frac{\pi}{\Delta\omega}\right) R(\theta_1). 
\end{equation}
This is a cascade of three pulses interrupted by two free evolutions of length $\pi/\Delta\omega$.
Note that the free evolution for a time $t=\pi/\Delta\omega$ leads to
an identity operation in the qubit subspace and a phase shift of $\pi$ on the leakage
transition, i.e. $\ket{0/1}\rightarrow \ket{0/1}$ and $\ket{L}\rightarrow-\ket{L}$.
Thus, in the isolated qubit subspace, we have just the three rotations with
the condition $2\theta_1 + \theta_2=\pi/2$ (e.g.~$\theta_1=\pi/8$ and $\theta_2=\pi/4$) for a NOT gate.
The dynamics on the leakage transition can be identified with time-dependent perturbation theory
after diagonalizing the qubit subspace in (\ref{eq:Hamiltonian}) to
lift the
degeneracy. This pulse sequence removes transitions to the third level
to first order in $\lambda/\Delta\omega$, with errors of the order $(\lambda/\Delta\omega)^2$ remaining.
As a comparison, the rotation with only one weak pulse discussed
in the previous paragraph introduces errors of the order
$\lambda/\Delta\omega$.
In principle, recursive application of such pulse sequences could
remove the error up to any desired order, but would lead to a
complicated and long pulse.
Also, it needs to be pointed out that
the rotations themselves require small $\lambda/\Delta\omega$ and
thus inevitably take significant extra time beyond the waiting periods.
In this work, we find optimal sequences with respect to fidelity and pulse duration.
We resort to numerical methods of optimal control, namely the
Gradient Ascent Pulse Engineering (GRAPE) algorithm \cite{Khaneja05}.

{\it Control theory} $-$ Given a Hamiltonian $H(t) = H(\lambda(t))$
such as Eq.~(\ref{eq:HamiltonianRwa}),
the goal is to find a function for the control parameter
$\lambda(t)$, $t\in[0,t_g]$,
such that a desired unitary quantum gate $U_{\rm F}$ is generated.
Formally, this can be achieved by minimizing the Euclidean distance between target
and actual evolution, i.e.~
$||U_{\rm F}- U(t_g)||_2^2 = ||U_{\rm F}||_2^2 + ||U(t_g)||_2^2 - 2\, {\rm Re}\, {\rm tr}
\left( U^{\dagger}_{\rm F} U(t_g) \right )$.
Here, $U(t)$ is the usual time evolution operator obeying the Schr\"odinger
equation $\dot{U}(t)=-\frac{{\rm i}}{\hbar}H(t)U(t)$ with initial condition $U(0)=\mathbbm{1}$.
Minimizing the Euclidean distance is, up to a global phase, equivalent to maximizing the fidelity
$\phi_1=\frac{1}{9} |{\rm tr}(U_{\rm F}^{\dagger}U(t_g))|^2$.
The fidelity $\phi_1$ is not sensitive to the global phase $\varphi_1$, as required above.
Additionally, we want to be insensitive to the relative phase of the leakage level.
Averaging over the free phase $\varphi_2$ and keeping only the relevant terms leads to,
\begin{equation}\label{eq:FidelitySubspace}
\phi_2=\frac{1}{4}\left(|\bra{0}U_{\rm F}^{\dagger}U(t_g)\ket{0}+\bra{1} U_{\rm F}^{\dagger}U(t_g)\ket{1} |^2 \right).
\end{equation}
In other words, it is sufficient to consider only the qubit subspace
when evaluating the gate performance $-$ everything else is fixed by
unitarity of $U(t_g)$: maximizing $\phi_2$ automatically eliminates transitions to
the third level. Optimization of fidelity $\phi_2$ requires the calculation of its gradient with respect
to the controls. To this end, we introduce the approximate time evolution
$U(t_g)\approx U_N U_{N-1}...U_1$. The time interval $[0, t_g]$ is sliced into N parts of length $\Delta t$,
on each of which the controls and therefore
the Hamiltonian is assumed to be constant. The propagator for an individual time step $U_j$ $(j=1...N)$
can thus be written as
$U_j={\rm exp}\left({-\frac{{\rm i}}{\hbar}\Delta t H(\lambda_j)}\right)$,
where $\lambda_j$ is the control amplitude during the $j$th time slice.
Along the lines of \cite{Khaneja05}, the gradient of the fidelity (\ref{eq:FidelitySubspace})
with respect to $\lambda_j$ is given by,
\begin{eqnarray}\label{eq:GradientSubspace}
\frac{\partial \phi_2}{\partial \lambda_j} &=&
-\frac{{\rm i}\Delta t}{2}\rm{Re} \left\{ \sum_{k=0,1}
\bra{k} U^{\dagger}_{\rm F} U_N...U_{j+1} \frac{\partial H}
{\partial \lambda_j}U_j...U_1 \ket{k} \right.\nonumber \\
 && \left.\sum_{m=0, 1} \bra{m} U^{\dagger}_{\rm F} U(t_g) \ket{m} \right\} .
\end{eqnarray}
This gradient is used in the GRAPE algorithm to find the maximum of the fidelity
Eq.~(\ref{eq:FidelitySubspace}) as a function of the control parameters $\lambda_j$.
Note that Eq.~(\ref{eq:GradientSubspace}) is valid when the time slices $\Delta t$
are small compared to the characteristic time scales of the system.

\begin{figure}
\includegraphics[scale=1.1]{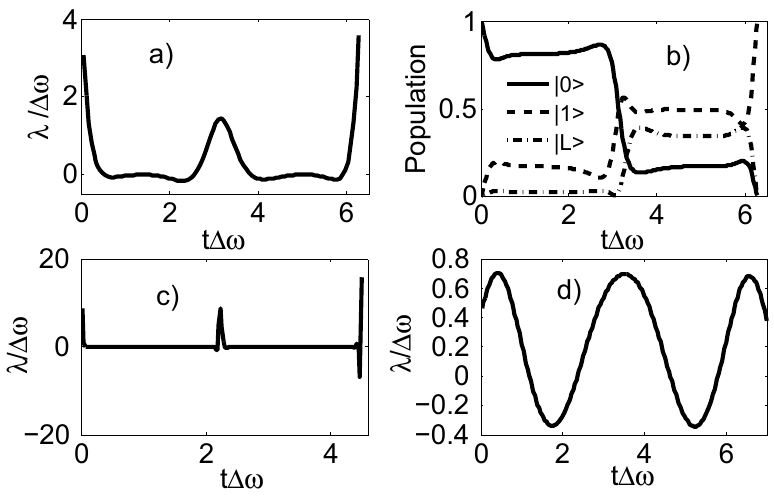}
\includegraphics[scale=1.0]{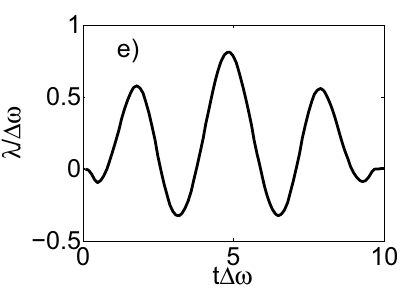}
\caption{
Control fields for a high-fidelity spin flip gate for a qubit in the presence of coherent transitions to a non-qubit level.
The control field with the duration $t_{\rm opt}=2\pi/\Delta\omega$, pictured in (a), is optimal in
the sense of a short gate time and high fidelity of $1-\phi_2 < 10^{-4}$. A cascade of three pulses is found
that harnesses the intrinsic resource of the system. In panel (b) the population of the three levels
during the application of that pulse is shown; the leakage level is populated during the pulse
and completely emptied in the end. In (c) and (d) a shorter ($t_g=4.5/\Delta\omega$) and longer ($t_g=7.0/\Delta\omega$)
pulse is depicted. These control fields are optimized without constraints to the pulse shape. Panel (e)
demonstrates a control field at $t_g=10.0/\Delta\omega$ with a smooth pulse rise, obtained with a penalty function
method. Nevertheless, a fidelity of $1-\phi_2 < 10^{-8}$ is achieved.
}
\label{fig:Pulses}
\end{figure}

{\it Pulses and performance} $-$ The initial guess for the optimization is the $\pi/2$-pulse  one would use in a two-level system, i.e.~$\lambda=\pi/(2t_g)$ for all $t\le t_g$. This pulse is then optimized using GRAPE based on the fidelity $\phi_2$.
For now, we assume that the controls $\lambda(t_i)$ can take arbitrary values
and that they can change arbitrary quickly from one time slice to the other.
We assume further that we can permit arbitrary excursions to $\ket{L}$
during the pulse.
As a result, at time $2\pi/\Delta\omega$ we recover a pulse similar to the approximate analytical solution,
see Fig.~\ref{fig:Pulses} (a). GRAPE finds a symmetric sequence of
three soft yet high amplitude, approximately Gaussian pulses
interrupted by two periods of free evolution. The areas under the three single pulse elements approximately
correspond to the angles discussed earlier. Deviations are compensated by the negative amplitude during the
free evolution parts.
Fig.~\ref{fig:Pulses} (b) shows the population of the qubit and the leakage level for the initial state being $\ket{0}$.
The first pulse of the cascade transfers about 20\% population to the excited state $\ket{1}$ while the leakage level
remains almost unpopulated. After the first wait period, the second pulse populates the leakage level around 40\%,
where it remains during the second wait period. Here, the leakage level accumulates a relative phase such that the third
pulse leads to complete depopulation.
For times below $t_{\rm opt}$, pulses such as in Fig.~\ref{fig:Pulses} (c) are obtained.
The pulse optimization tries to improve fidelities by rapidly turning the control field to very high amplitudes.
However, the integral resource of the $2\pi/\Delta\omega$ waiting time is not available,
which lowers the attainable fidelities.
For longer gate times than $t_{\rm opt}=2\pi/\Delta\omega$ the optimal control field is a smooth pulse shape, see Fig.~\ref{fig:Pulses} (d).
As an additional benefit, the control amplitudes are lower, leading to less transition to the non-qubit level during
the application of the pulse.

\begin{figure}
\includegraphics[scale=0.4]{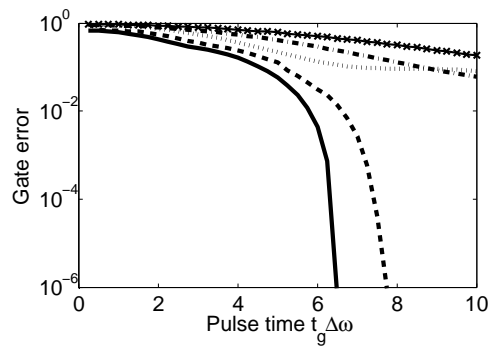}
\caption{
The attainable gate error of a spin flip gate using various control strategies is demonstrated as a function
of pulse duration. 
Rectangular ($\cdot \cdot \cdot$) and Gaussian pulses with $\alpha=2$ ($- \cdot$) and $\alpha=3$ ($\times$)
only achieve limited fidelities at short pulse durations.
GRAPE optimized control fields ($-$) consistently improve the gate error, with $1-\phi_2 < 10^{-8}$
after an optimal gate time of $t_{\rm opt}=2\pi/\Delta\omega$.
Pulses that have a smooth initial rise slightly increase the gate error and shift the optimal gate time ($- -$).
The pulse rise times are around $1.0/\Delta\omega$, which is obtained by choosing the penalty
function (\ref{eq:PenaltyStrength}) with $t_0=0.1/\Delta\omega$ and $\gamma=5.0/\Delta\omega$.
}\label{fig:FidelityVersusTime}
\end{figure}

Fig.~\ref{fig:FidelityVersusTime} shows the gate error $1-\phi_2$ versus pulse duration
$t_g$ for unoptimized and optimized pulses.
We see that GRAPE pulses easily achieve errors of $1-\phi_2=10^{-6}$ and below.
It has been estimated by various authors assuming different error models that
the threshold error rate for fault-tolerant quantum computing
lies between $10^{-2}$ and $10^{-6}$ \cite{Knill, Steane03, NielsenBook, Aharonov}.
Within the present model, pulse shaping easily reaches these thresholds for gate durations greater $t_{\rm opt}$.
Optimizing fidelity $\phi_2$ is advantageous over optimizing fidelity $\phi_1$:
the irrelevant relative phase of the leakage level is not taken care of and therefore
higher gate fidelities can be obtained.
For comparison, we also show the gate error of standard pulse shapes like Gaussian and rectangular pulses \cite{Steffen03}.
Gaussian pulses used here are given by $\lambda(t) = \alpha/t_g \sqrt{\pi/2}\ {\rm exp}(-\alpha^2/t_g^2 (t-t_g/2)^2)$.
These pulse shapes have attainable gate errors of $10^{-1}-10^{-2}$ at the pulse durations considered here,
orders of magnitudes worse than the optimized pulses. For rectangular pulses, only at gate times of $t_g > 280/\Delta\omega$ errors below
$10^{-4}$ can be achieved. Essentially, these long gate times mean low control amplitudes and thus only small population
of the leakage level: the weak-driving limit where $\lambda/\Delta\omega \ll 1$.

{\it Improving the pulse shape} $-$ The pulses presented so far are obtained by solely optimizing the fidelity $\phi_2$,
Eq.~(\ref{eq:FidelitySubspace}).
As seen in Fig.~\ref{fig:Pulses} the controls have sharp initial rises, which may pose problems in the experimental implementation.
In this section, we show that controls with smooth rises can be obtained at a small cost of pulse duration.
We use the concept of penalty functions to constrain the gradient search algorithm.
High amplitudes at the beginning and the end of the pulse can be penalized
by amending the fidelity (\ref{eq:FidelitySubspace}),
\begin{equation} \label{eq:FidelityPenaltyRise}
\tilde{\phi}_2=\phi_2 - \int_0^{t_g}\gamma_{\rm A}(t)\lambda^2(t)dt.
\end{equation}
Here, we take the penalty strength as a function of time to be of the form
\begin{equation}\label{eq:PenaltyStrength}
\gamma_{\rm A}(t) = \gamma \left( 2-\tanh\left(\frac{t}{t_0}\right) + \tanh\left(\frac{t_g-t}{t_0}\right) \right),
\end{equation}
where the positive $\gamma$ is the overall strength of the penalty and
$t_0$ essentially parameterizes the rise time of the pulse.
A non-zero $\gamma_{\rm A}(t)\lambda^2(t)$ for any time t will reduce the fidelity.
A smooth penalty like in Eq.~(\ref{eq:PenaltyStrength}) leads to smooth pulses shapes.
The gradient of $\tilde{\phi}_2$ is used in the optimization procedure.

Fig.~\ref{fig:Pulses} (e) shows the envelope of a pulse with $t_g=10.0/\Delta\omega$,
obtained by optimizing the fidelity (\ref{eq:FidelityPenaltyRise}) and (\ref{eq:PenaltyStrength})
with the parameters $\gamma=5.0/\Delta\omega$ and $t_0=0.1/\Delta\omega$.
The control field starts at zero and increases within a time
of around $0.5/\Delta\omega$ to the first peak of around $-0.1\Delta\omega$ and
within a time of around $1.8/\Delta\omega$ to the second peak of around $0.6\Delta\omega$.
This pulse has a gate error below $10^{-8}$.
In Fig.~\ref{fig:FidelityVersusTime}, the gate error of the amplitude-constrained pulses
is demonstrated as a function of the pulse time. The parameters are again $\gamma=5.0/\Delta\omega$ and $t_0=0.1/\Delta\omega$.
The introduction of a penalty for the controls comes at a small cost
of fidelity. The optimal gate time $t_{\rm opt}$ of the unconstrained case is shifted to around $7.75/\Delta\omega$.
After that gate time, errors of below $10^{-6}$ are obtained, due to the fact that the $2\pi/\Delta\omega$
free evolution and the penalty requirements can be easily incorporated into a shaped control field.

{\it Discussion of the time scales} $-$
We discuss the control fields and their properties in terms of the actual potential anharmonicity
observed in the latest experiments in phase qubits and the transmon. In the phase qubit, the detuning
is given in Fig.~5 of the supplementary material in \cite{LuceroArxiv} to be $\Delta\omega/2\pi= 0.2\ $GHz.
Thus, the optimal gate time is $t_{\rm opt}= 5\ $ns.
This compares favorably to the $8$ns FWHM Gaussian pulse employed in \cite{LuceroArxiv} with experimental gate fidelity of 0.98.
The amplitudes of the pulse sequence at $t_{\rm opt}$, Fig.~\ref{fig:Pulses} (a), are around $\lambda/2\pi=0.8\ $GHz,
while the FWHM of the three approximately (half) Gaussian pulses is less that $1$ns.
Other pulses, e.g.~Fig.~\ref{fig:Pulses} (e) have a Rabi frequency and a modulation frequency of the
order of the detuning.
In the transmon, the detuning is slightly larger, $\Delta\omega/2\pi=0.455\ $GHz \cite{Schreier08},
leading to an optimal gate time of $t_{\rm opt}=2.2\ $ns.

{\it Conclusion} $-$ We have shown that high-fidelity quantum gates in a qubit can be performed despite the
presence of a non-qubit leakage level.
Our model approximates the resonantly driven Josephson phase qubit and the transmon, and
is with modifications applicable to optical lattices.
We have elucidated a composite sequence of pulses that performs a spin flip gate in
the qubit subspace and refocusses the leakage transition to first order in the weak driving limit.
Numerical optimization of the control field with the GRAPE algorithm drastically improves fidelities and pulse durations,
leading to smooth, low-amplitude pulse shapes. We have identified an optimal pulse time $t_g\ge 2\pi/\Delta\omega$
that is integral to the problem and above which gate errors for a spin flip operation are below $1-\phi_2<10^{-8}$.
As a comparison, the rectangular pulse would take $t_g\approx280/\Delta\omega$ for a gate error of $10^{-4}$.
We have modified the pulse optimization with a penalty function such that the control fields become easier to realize in
the experimental implementation. We have shown that this constraint can lead to similar near-perfect gate errors
when the pulse duration is lengthened by about twice the required rise time.

We would like to acknowledge useful discussions with A.G.~Fowler, J.~Gambetta, J.M.~Martinis,
R.~McDermott, F.~Motzoi, and T.~Schulte-Herbr\"uggen. This work was supported
by NSERC through a discovery grant and Quantumworks.

\end{document}